\documentclass[aps,twocolumn,showpacs]{revtex4}
\usepackage{amssymb}
\usepackage{graphicx}
\usepackage{amsmath}
\usepackage{amsfonts}

\begin{document}

\title{Symmetry of superconducting states with two orbitals on a tetragonal
lattice: application to $LaFeAsO_{1-x}F_{x}$}
\date{\today}
\author{Yi Zhou$^{1,2}$, Wei-Qiang Chen$^1$ and Fu-Chun Zhang$^1$ }
\affiliation{Department of Physics, Center of Theoretical and Computational Physics, The
University of Hong Kong, Hong Kong, China \\
Department of Physics, Chinese University of Hong Kong, Hong Kong, China}

\begin{abstract}
We use group theory to classify the superconducting states of systems with
two orbitals on a tetragonal lattice. The orbital part of the
superconducting gap function can be either symmetric or anti-symmetric. For
the orbital symmetric state, the parity is even for spin singlet and odd for
spin triplet; for the orbital anti-symmetric state, the parity is odd for
spin singlet and even for spin triplet. The gap basis functions are obtained
with the use of the group chain scheme by taking into account the spin-orbit
coupling. In the weak pairing limit, the orbital anti-symmetric state is
only stable for the degenerate orbitals. Possible application to iron-based 
superconductivity is discussed.
\end{abstract}
\maketitle

\section{Introduction}

Symmetry plays an important role in the study of superconductivity. By using the
symmetry of the superconducting (SC) gap function, Ginzburg-Landau theory
can be constructed and electromagnetic response and topological excitations
can be inspected. In the past decades, the symmetry analyses to classify
unconventional SC states have been focused on single-band superconductors,
and have shed much light on our understanding of heavy-fermion and ruthenate
superconductors\cite{Sigrist91}.

Very recently, a new class of iron-based high temperature superconductors
has been discovered with $T_{c}$ as high as above $50$K\cite{LOFS,COFS,
SOFS,POFS,SOFS2,GOFS,SOxFS2,Hole}. Experimentally, spin density wave (SDW)
order has been observed in the parent compound $LaFeAsO$, but vanishes upon
fluorine doping where the superconductivity appears\cite%
{SDW,Neutron}. Specific heat measurement as well as nuclear
magnetic resonance suggested line nodes of the SC gap\cite{CV,NMR,NMR2,NMR3}%
. The transition temperature estimated based on the electron-phonon coupling
is low, and unlikely to explain the observed superconductivity\cite{Boeri}.
It has been proposed that the superconductivity is of magnetic origin and is
unconventional. Local density approximation (LDA) shows that iron's $3d$
electrons dominate the density of states near the Fermi surfaces in the
parent compound $LaFeAsO$\cite%
{LDA1Singh,LDA2Kotliar,LDA3Xu,LDA4Mazin,LDA5Cao,LDA6Ma}. In their
calculations, there are three hole-like Fermi surfaces centered at the $%
\Gamma $ point and two electron-like Fermi surfaces around the $M$ point. By
F-doping, the area of the three hole-like Fermi surfaces shrinks while the
area of the two electron-like Fermi surfaces expands. The band structure
obtained from the LDA may be well modeled by a tight-binding model with two or three
orbitals ($d_{xz}$, $d_{yz}$ and $d_{xy})$ \cite%
{Theory1Dai,Theory2LiTao,Theory3Xiaogang,Theory4ZYWeng,Theory5Shoucheng,Theory6Zidan, Theory7Qimiao,Theory8Scalapino}%
. Because of the multiple orbitals in the low energy physics, it is natural
to raise the question how to generalize the symmetry consideration from
single-band to multi-band cases.

In this paper, we will generalize the symmetry analyses developed for the
single band SC state to systems with two orbitals. We will use group theory
to classify the allowed symmetry of the gap functions of the two-orbital SC
state on a tetragonal lattice by including a spin-orbit coupling between the
paired electrons. While our focus will be on the Fe-based compounds, some of
our analyses may be applied to more general systems with two orbitals.

We arrange this paper as the follows. In Sec. II, we discuss the
symmetries governing the system and how these symmetries affect the
Hamiltonian and gap functions. In Sec. III, we consider the possible
two-orbital SC states on a tetragonal lattice. Section IV is devoted to
summary and discussions. We also supply some appendixes for details. In
Appendix \ref{xi}, we show how the symmetries give rise to the requirements
to the non-interacting Hamiltonian. In Appendix \ref{gD4h}, we specify the
point group $D_{4h}$ of lattice according to space group $P4/nmm$. In
Appendix \ref{transfer}, we discuss how the gap functions transfer under
symmetry operations. In Appendix \ref{gap}, we discuss the energy gap
functions in the degenerate bands.

\section{Symmetry of gap function $\Delta \left( \mathbf{k}\right) $}

We consider a tetragonal lattice, appropriate for doped $LaFeAsO$. Since our
primary interest is in the SC state, we will not consider the translational
symmetry broken state such as the spin density wave state observed in the
parent compound of $LaFeAsO$. The system is invariant under both time
reversal and space inversion. The inversion symmetry suggests that the SC
pairing is either even or odd in parity. We shall assume in this paper that
the time reversal symmetry remains unbroken.

We consider a system described by Hamiltonian 
\begin{equation}
H=H_{0}+H_{pair}+H_{so}  \label{H}
\end{equation}%
where $H_{0}$ is non-interacting part, and $H_{pair}$ is a pairing
Hamiltonian, and $H_{so}$ is the spin-orbit coupling of the Cooper pairs.
We shall consider the SC state preserves all the symmetries in $H_{0}$
except the $U(1)$ symmetry in electric charge and the spin rotational
symmetry due to a weak $H_{so}$. We assume $H_{0}$ to be given by a
tight-binding Hamiltonian 
\begin{equation}
H_{0}=\sum\limits_{\mathbf{k}\alpha _{1}\alpha _{2}s}c_{\mathbf{k}\alpha
_{1}s}^{\dagger }\xi _{\mathbf{k\alpha }_{1}\alpha _{2}}c_{\mathbf{k}\alpha
_{2}s},  \label{H0}
\end{equation}%
where $\alpha =1,2$ are the orbital indices, which correspond to the two
orbitals $3d_{xz}$ and $3d_{yz}$ in $Fe$, $s=\uparrow ,\downarrow $ are the
spin indices. Note that for $LaFeAsO$, the actual crystal structure has two
Fe-atoms in a unit cell due to the $As$ atomic positions, which are
allocated above and below the Fe-plane alternatively. For convenience, here
we use the extended Brillouine zone, and the summation $\mathbf{k}$ is in
the extended zone. $H_{0}$ is invariant under symmetry transformation. This
requires certain symmetries on $\xi _{\mathbf{k\alpha }_{1}\alpha _{2}}$,
which we will discuss in detail in Appendix \ref{xi}. We note that a more appropriate 
model should also include the $d_{xy}$-orbital\cite{Theory3Xiaogang}, 
but we shall leave the symmetry analyses of the three orbitals for future study, and 
consider a simplified version of the two orbital case in this paper.

\begin{figure}[htpb]
\includegraphics[width=8.0cm]{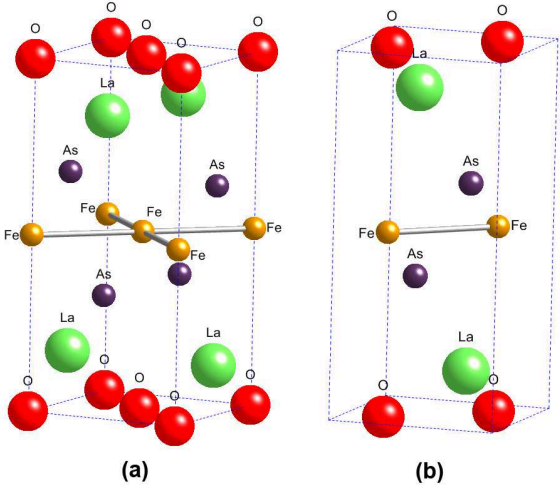}
\caption{(color online) Lattice structure of $LaFeAsO$. It is a tetragonal lattice with two 
$Fe$ atoms per unit cell. The lattice constants are $a=b\simeq4.03\mathring{A%
}$ and $c\simeq8.74\mathring{A}$\cite{Neutron}, where $a$ is the
distance between two next nearest neighbor $Fe$ atoms. (a) Origin choice 1
of space group $P4/nmm$, at $\bar{4}m2$ and at $(-a/4,a/4,0)$ from center ($%
2/m$). It can be chosen either at an $Fe$ or at an $O$ atom; (b) Origin
choice 2 of space group $P4/nmm$, at center ($2/m$) and at $(a/4,-a/4,0)$
from $\bar{4}m2$. It can be chosen either at the midpoint of two nearest
neighbor $Fe$ atoms or at the midpoint of two nearest neighbor $O $ atoms.
Here $2/m$ denotes the two fold rotation $C_{2} $ and reflection $m$ (see
Appendix \ref{gD4h} for details). The origin choice 1 and 2 are
different from each other by a shift of $(-a/4,a/4,0)$\cite{crystallography}.}
\label{lat}
\end{figure}

The gap function of the two-orbital SC state can be generally written as 
\begin{equation}
\Delta _{s_{1}s_{2}}^{\alpha _{1}\alpha _{2}}\left( \mathbf{k}\right)
=-\sum\limits_{\substack{ \mathbf{k}^{\prime }\alpha _{3}\alpha _{4}  \\ %
s_{3}s_{4}}}V_{s_{2}s_{1}s_{3}s_{4}}^{\alpha _{2}\alpha _{1}\alpha
_{3}\alpha _{4}}\left( \mathbf{k},\mathbf{k}^{\prime }\right) \langle c_{%
\mathbf{k}^{\prime }\alpha _{3}s_{3}}c_{-\mathbf{k}^{\prime }\alpha
_{4}s_{4}}\rangle ,  \label{delta}
\end{equation}%
where $V_{s_{2}s_{1}s_{3}s_{4}}^{\alpha _{2}\alpha _{1}\alpha _{3}\alpha
_{4}}\left( \mathbf{k},\mathbf{k}^{\prime }\right) $ is the effective
attractive interaction. Hereafter we will use the matrix notation $\Delta
\left( \mathbf{k}\right) $ for the gap function.

To classify the symmetry of the SC gap function for multiple orbitals, we
recall that in the single orbital case, the spin-orbit coupling of the
Cooper pair plays an important role to the non s-wave superconductors, and
the symmetry of the gap function is determined by the crystal point group of
the lattice and the spin part of the gap function. In the two-orbital
system, the orbital degree of freedom is usually coupled to the crystal
momentum, hence to the spin via the spin-orbit coupling. Therefore, the
spin, spatial, and the orbital parts are generally all related in the gap
function.

Let us first discuss the crystal symmetry. The crystal structure of $LaFeAsO$
is shown in Fig. 1. The tetragonal crystal symmetry is characterized by the
point group $D_{4h}$. The tetragonal point group may be specified according
to the space group $P4/nmm$ of the compound, and the details will be
discussed in Appendix \ref{gD4h}. There are five irreducible representations
of $D_{4}$ group, denoted by $\Gamma $, including 4 one-dimensional
representations ($A_{1}$, $A_{2}$, $B_{1}$ and $B_{2}$) and 1
two-dimensional representation ($E$)\cite{Butler}. The tetragonal lattice
symmetry requires $H_{0}$ to be a \textquotedblleft
scalar\textquotedblright\ or $A_{1}$ representation of $D_{4}$. In the
absence of spin-orbit coupling, spin is rotational invariant and we have
both the point group symmetry and the spin rotational symmetry.\cite{P4nmm}

We now discuss the orbital degrees of freedom in connection with the crystal
symmetry. The two orbitals $d_{xz}$ and $d_{yz}$ transform as $E$
representation of $D_{4}$. In general the orbital indices $d_{xz}$ and $%
d_{yz}$ are not good quantum numbers because of the mixed term of the two
orbitals in $H_{0}$, and the two energy bands are not degenerate. In that
case it is necessary to include the coupling of the orbital to spatial and
spin degrees of freedom.

Without loss of generality, the gap function can be written as a linear
combination of the direct products of the orbital part $\Omega $ and the
spin part $\Delta ^{spin}$ in a given representation $\Gamma $ of the point
group $D_{4}$, 
\begin{eqnarray}
\Delta \left( \Gamma ;\mathbf{k}\right) &=&\sum\limits_{m,\Gamma
_{LS},\Gamma _{\Omega }}\eta \left( \Gamma ,m\right) \left\langle \Gamma
,m|\Gamma _{LS},m_{LS};\Gamma _{\Omega },m_{\Omega }\right\rangle  \notag \\
&&\times \Delta ^{spin}\left( \Gamma _{LS},m_{LS};\mathbf{k}\right) \otimes
\Omega \left( \Gamma _{\Omega },m_{\Omega }\right) ,  \label{delta1}
\end{eqnarray}%
where both $\Delta ^{spin}$ and $\Omega $ are $2\times 2$ matrices, $\Delta
_{s_{1}s_{2}}^{spin}$ dictates the pairing in spin space and $\Omega
_{\alpha _{1}\alpha _{2}}$ dictates the pairing in orbital space, $\Gamma
_{LS}$ and $\Gamma _{\Omega }$ are irreducible representations of $D_{4}$ in
spin and orbital spaces, respectively, $m,m_{LS},m_{\Omega }$ are bases of
representations $\Gamma ,\Gamma _{LS},\Gamma _{\Omega }$, respectively. $%
\left\langle \Gamma ,m|\Gamma _{LS},m_{LS};\Gamma _{\Omega },m_{\Omega
}\right\rangle $ is the Clebsch-Gordan (CG) coefficient. Note that the $%
\mathbf{k}$-dependence is contained in $\Delta ^{spin}$, but not in $\Omega $%
. Here $\eta \left( \Gamma ,m\right) $ is the coefficient of the basis $m$
of the representation $\Gamma $. The anti-symmetric statistics of two
electrons requires 
\begin{equation}
\Delta ^{T}\left( -\mathbf{k}\right) =-\Delta \left( \mathbf{k}\right) .
\label{fasym}
\end{equation}

Below we will first discuss $\Delta ^{spin}\left( \Gamma _{LS};\mathbf{k}%
\right) $ and $\Omega \left( \Gamma _{\Omega }\right) $ separately, and then
combine the two to form an irreducible representation $\Gamma $ of $D_{4}$.
We follow Sigrist and Ueda\cite{Sigrist91} and write $\Delta ^{spin}\left( 
\mathbf{k}\right) $ in terms of the basis functions $\psi \left( \Gamma ,m;%
\mathbf{k}\right) $ for the spin singlet $S=0$ and $\mathbf{d}\left( \Gamma
,m;\mathbf{k}\right) $ for the spin triplet $S=1$, 
\begin{equation}
\Delta ^{spin}\left( \Gamma ,m;\mathbf{k}\right) =i\left[ \sigma _{0}\psi
\left( \Gamma ,m;\mathbf{k}\right) +\mathbf{\sigma }\cdot \mathbf{d}\left(
\Gamma ,m;\mathbf{k}\right) \right] \sigma _{2},  \label{deltas}
\end{equation}%
Here $\psi \left( \mathbf{k}\right) $ is a scalar and $\mathbf{d}\left( 
\mathbf{k}\right) $ is a vector under the transformation of spin rotation.
For this reason, it is more convenient to use $\psi \left( \mathbf{k}\right) 
$ and $\mathbf{d}\left( \mathbf{k}\right) $ instead of $\Delta ^{spin}\left( 
\mathbf{k}\right) $ to classify the pairing states.

Due to the fermionic anti-symmetric nature, the gap function must be
anti-symmetric under the two particle interchange, or under a combined
operations of space inversion, interchange of the spin indices and
interchange of the orbital indices of the two particles. Let $P_{1,2}$ be
the two particle interchange operator, and $P_{space},P_{spin},P_{orbital}$
be the interchange operator acting on the space, spin, and orbital,
respectively, then the fermion statistics requires 
\begin{equation}
P_{1,2}=P_{space}P_{spin}P_{orbital}=-1.
\end{equation}%
Since the system is of inversion symmetry, the pairing states must have
either even parity $P_{space}=+1$ or odd parity $P_{space}=-1$. Furthermore,
the total spin $S$ of the Cooper pair is a good number, and this is so even
in the presence of $H_{so}$, which breaks spin rotational symmetry but
keeps inversion symmetry, so that it does not mix the $S=1$ with $S=0$ states.
Therefore, under the two particle interchange, the spin part of the gap
function must be either symmetric: ($P_{spin}=+1$, with $S=1$) or
anti-symmetric ($P_{spin}=-1$ with $S=0$), represented by the vector $%
\mathbf{d}$ or the scalar $\psi $ in Eq. (\ref{deltas}), respectively. Because of the
inversion and spin symmetries, we have $P_{orbital}=\pm 1$.

The orbital part of the pairing matrix $\Omega $ is spanned in the vector
space of $\left( d_{xz},d_{yz}\right) $, which is an irreducible
representation $E$ of the point group $D_{4}$. Thus $\Omega $ belongs to an
irreducible representation given by $E\otimes E=A_{1}\oplus A_{2}\oplus
B_{1}\oplus B_{2}$, which are all one-dimensional, hence simplifies the
classification of the pairing states. According to the CG coefficients of $%
D_{4}$ group, up to a global factor, $\Omega =\sigma _{0}$ in representation 
$A_{1}$, $\Omega =\sigma _{3}$ in $B_{1}$, and $\Omega =\sigma _{1}$ in $%
B_{2}$, which are all orbital symmetric: $P_{orbital}=+1$. $\Omega =\sigma
_{2}$ in $A_{2}$ representation, which is orbital anti-symmetric: $%
P_{orbital}=-1$. In brief, $A_{1}$ and $B_{1}$ of $\Omega $ are
representations for intra-orbital pairing, $B_{2}$ is for symmetric
inter-orbital pairing and $A_{2}$ is for anti-symmetric inter-orbital
pairing. For convenience, we choose $\Omega $ to be Hermitian, so that $\psi
\left( \mathbf{k}\right) $ and $\mathbf{d}\left( \mathbf{k}\right) $ will be
real. (Wan and Wang\cite{QHWang} pointed out that Pauli matrices transfer as four
one-dimensional irreducible representations.)

The crystal point group of the lattice will dictate the allowed symmetry in $%
\mathbf{k}$ space. The transformation of $\psi $ and $\mathbf{d}$ under
symmetry operations can be found in Appendix \ref{transfer}. In Sec. III, 
we will study the basis functions $\psi \left( \Gamma ,m;\mathbf{k}%
\right) $ and $\mathbf{d}\left( \Gamma ,m;\mathbf{k}\right) $, and combine
them with the orbital part $\Omega $ to obtain the irreducible
representations of group $D_{4}$.

\section{Possible two-orbital SC states on a tetragonal lattice}

We will use the group chain scheme to study the representation and the basis
function of $\psi $ and $\mathbf{d}$ by assuming a spin-orbit coupling. In
the group chain scheme, we begin with a rotational invariant system in both
spin and spatial spaces. The representation of its symmetry group $G$ can be
decoupled into a spatial part $D_{\left( L\right) }$ and a spin part $%
D_{\left( S\right) }$, with $\mathbf{L}$ as the relative angular momentum of
the Cooper pair, 
\begin{equation}
D_{\left( G\right) }=D_{\left( L\right) }\otimes D_{\left( S\right) },
\end{equation}%
In the presence of the spin-orbit coupling, $D_{\left( L\right) }$ and $%
D_{\left( S\right) }$ are no longer the irreducible representation of the
rotational group, but the total angular momentum $\mathbf{J}=\mathbf{L}+%
\mathbf{S}$ is, and $D_{\left( J\right) }$ is the corresponding irreducible
representation of the rotational group.

We now turn on a crystal field with tetragonal lattice symmetry group $D_4 $%
, so that the rotational group $SO(3)$ is reduced to $D_4$, and $D_{\left(
L\right) }\otimes D_{\left( S\right) }$ is reduced to a direct product of
irreducible representations $\Gamma_{LS}$ of group $D_{4}$, 
\begin{equation}
D_{\left( L\right) }\otimes D_{\left( S\right) } \rightarrow
\bigoplus\limits_{\Gamma _{LS}}D_{\left( \Gamma _{LS}\right) }.
\end{equation}
Including the coupling to the orbital part $\Omega$, the representation $%
D_{\left( \Gamma _{LS}\right) }\otimes D_{\left( \Gamma _{\Omega }\right) }$
is decomposed into irreducible representations, 
\begin{equation}
D_{\left( \Gamma _{LS}\right) }\otimes D_{\left( \Gamma _{\Omega }\right)
}=\bigoplus\limits_{\Gamma }D_{\left( \Gamma \right) }.
\end{equation}%
$D_{\left( \Gamma _{\Omega }\right) }$ is one-dimensional, thus these
representations have a very simple form.

Let us consider the even parity case. From Eq. (7), the SC gap function can
be either orbital symmetric $P_{orbital}=+1$, spin singlet or orbital
anti-symmetric $P_{orbital}=-1$, spin triplet. We list the SC gap basis
functions for spin singlet and spin triplet according to the irreducible
representations $\Gamma $ in Tables I and II\ respectively. The listed even
pairing states include $s$-wave (extended $s$-wave), $d$-wave and $g$-wave.
Here $0$, $\tilde{0}$, $2$, $\tilde{2}$, and $1$ are natural notation for
the five irreducible representations of $D_{4h}$; $A_{1}$, $A_{2}$, $B_{1}$, 
$B_{2}$ and $E$ are Sch\"{o}nflies notation; $\Gamma _{1-5}$ are Koster
notation. According to Eq. (\ref{delta}), the gap function of the SC state
is a linear combination of the basis functions in one irreducible
representation $\Gamma $, and the basis functions belonging to different
representations in $\Gamma $, e.g. $A_{1g}$ and $B_{2g}$, will not mix with
each other.

We are particularly interested in 2D or quasi-2D limiting
cases, relevant to Fe-based SC compounds, where the gap function is $k_{z}$%
-independent, and the Fermi surface is cylinder-like. However, for
completeness we also list in the Tables those three-dimensional basic functions
marked with 3D.

In the last column of each table, we list the allowed energy zeroes in the
quasiparticle dispersion determined by the gap functions for the special
case that the two energy bands are completely degenerate. The detailed
calculations for the quasiparticle energies in the degenerate cases are
given in Appendix \ref{gap}. We will discuss the quasiparticle properties
for the non-degenerate cases in the discussion section below.

\begin{table*}[htbp]
\caption{Superconducting gap basis functions $\psi (\mathbf{k})$ on
tetragonal lattice for even parity, orbital symmetric and spin singlet
pairing states. $\Gamma$: representation of $D_4$. The listed notations are
natural, or Sch\"{o}nflies and Koster (in parentheses). $\Omega$: orbital
representation, $\sigma_0$ is the identity matrix, and $\sigma_{1,2,3}$ 
are Pauli matrices. Listed gaps properties are for the two
completely degenerate orbitals. $k_z$-dependent basis functions are marked
with $(3D)$, listed for completeness.}
\label{rpD1}%
\begin{tabular}{|c|c|c|c|}
\hline
$\Gamma $ & basis $\psi (\mathbf{k})$ & $\Omega $ & gap \\ \hline\hline
& $1,k_{x}^{2}+k_{y}^{2};k_{z}^{2}$ (3D) & $\sigma _{0}$ &  \\ \cline{2-3}
$0$ ($A_{1g}$, $\Gamma _{1}^{+}$) & $k_{x}^{2}-k_{y}^{2}$ & $\sigma _{3}$ & 
line nodal, or full gap \\ \cline{2-3}
& $k_{x}k_{y}$ & $\sigma _{1}$ &  \\ \hline\hline
& $k_{x}k_{y}(k_{x}^{2}-k_{y}^{2})$ & $\sigma _{0}$ &  \\ \cline{2-3}
$\tilde{0}$ ($A_{2g}$, $\Gamma _{2}^{+}$) & $k_{x}k_{y}$ & $\sigma _{3}$ & 
line,full \\ \cline{2-3}
& $k_{x}^{2}-k_{y}^{2}$ & $\sigma _{1}$ &  \\ \hline\hline
& $k_{x}^{2}-k_{y}^{2}$ & $\sigma _{0}$ &  \\ \cline{2-3}
$2$ ($B_{1g}$, $\Gamma _{3}^{+}$) & $1,k_{x}^{2}+k_{y}^{2};k_{z}^{2}$ (3D) & 
$\sigma _{3}$ & line,full \\ \cline{2-3}
& $k_{x}k_{y}(k_{x}^{2}-k_{y}^{2})$ & $\sigma _{1}$ &  \\ \hline\hline
& $k_{x}k_{y}$ & $\sigma _{0}$ &  \\ \cline{2-3}
$\tilde{2}$ ($B_{2g}$, $\Gamma _{4}^{+}$) & $k_{x}k_{y}(k_{x}^{2}-k_{y}^{2})$
& $\sigma _{3}$ & line,full \\ \cline{2-3}
& $1,k_{x}^{2}+k_{y}^{2};k_{z}^{2}$ (3D) & $\sigma _{1}$ &  \\ \hline\hline
$1$ ($E_{g}$, $\Gamma _{5}^{+}$) & $( k_{x}k_{z},k_{y}k_{z}) $ (3D) & $%
\sigma _{0},\sigma _{3},\sigma _{1}$ &  \\ \hline
\end{tabular}%
\end{table*}

\begin{table*}[htbp]
\caption{Superconducting gap basis functions $\mathbf{d}(\mathbf{k})$ on
tetragonal lattice for even parity, orbital anti-symmetric and spin triplet
pairing states. Notations are the same as in Table I. }
\label{rpD2}%
\begin{tabular}{|c|c|c|c|}
\hline
$\Gamma $ & basis $\mathbf{d}(\mathbf{k})$ & $\Omega $ & gap \\ \hline\hline
$0$ ($A_{1g}$, $\Gamma _{1}^{+}$) & $\hat{z},(k_{x}^{2}+k_{y}^{2})\hat{z}%
,(k_{x}^{4}+k_{y}^{4})\hat{z},k_{x}^{2}k_{y}^{2}\hat{z}$ & $\sigma _{2}$ & 
line,full \\ \hline\hline
$\tilde{0}$ ($A_{2g}$, $\Gamma _{2}^{+}$) & $k_{z}(k_{x}\hat{y}-k_{y}\hat{x}%
) $ (3D) & $\sigma _{2}$ &  \\ \hline\hline
$2$ ($B_{1g}$, $\Gamma _{3}^{+}$) & $(k_{x}^{2}-k_{y}^{2})\hat{z};k_{z}(k_{x}%
\hat{x}-k_{y}\hat{y})$ (3D) & $\sigma _{2}$ & line \\ \hline\hline
$\tilde{2}$ ($B_{2g}$, $\Gamma _{4}^{+}$) & $k_{x}k_{y}\hat{z};k_{z}(k_{x}%
\hat{x}+k_{y}\hat{y})$ (3D) & $\sigma _{2}$ & line \\ \hline\hline
$1$ ($E_{g}$, $\Gamma _{5}^{+}$) & $%
\begin{array}{c}
( \hat{x},\hat{y}) ,( k_{x}^{2}\hat{x},k_{x}^{2}\hat{y}) ,( k_{y}^{2}\hat{x}%
,k_{y}^{2}\hat{y}) ,( k_{x}k_{y}\hat{x},k_{x}k_{y}\hat{y}) ; \\ 
( k_{z}^{2}\hat{x},k_{z}^{2}\hat{y}) ,( k_{x}k_{z}\hat{z},k_{y}k_{z}\hat{z}) 
\text{ (3D)}%
\end{array}
$ & $\sigma _{2}$ & line,full \\ \hline
\end{tabular}%
\end{table*}

Similarly, for the odd parity pairing $P_{space}=-1$, we can have either
orbital anti-symmetric $P_{orbital}=-1$, spin singlet, or orbital symmetric $%
P_{orbital}=1$, spin triplet, which are listed in Tables III and IV
respectively. For the spin triplet, we list $p$-wave, $f$-wave and $h$-wave
states.

\begin{table*}[htbp]
\caption{Superconducting gap basis functions $\psi (\mathbf{k})$ on
tetragonal lattice for odd parity, orbital anti-symmetric and spin singlet
pairing state. Notations are the same as in Table I.}
\label{rpD3}%
\begin{tabular}{|c|c|c|c|}
\hline
$\Gamma $ & basis $\psi (\mathbf{k})$ & $\Omega $ & gap \\ \hline\hline
$0$ ($A_{1u}$, $\Gamma _{1}^{-}$) & $k_{z}$ (3D) & $\sigma _{2}$ &  \\ 
\hline\hline
$\tilde{0}$ ($A_{2u}$, $\Gamma _{2}^{-}$) & $%
k_{z}(k_{x}^{4}-6k_{x}^{2}k_{y}^{2}+k_{y}^{4})$ (3D) & $\sigma _{2}$ &  \\ 
\hline\hline
$2$ ($B_{1u}$, $\Gamma _{3}^{-}$) & $k_{z}(k_{x}^{2}-k_{y}^{2})$ (3D) & $%
\sigma _{2}$ &  \\ \hline\hline
$\tilde{2}$ ($B_{2u}$, $\Gamma _{4}^{-}$) & $k_{x}k_{y}k_{z}$ (3D) & $\sigma
_{2}$ &  \\ \hline\hline
$1$ ($E_{u}$, $\Gamma _{5}^{-}$) & $(k_{x},k_{y}) $ & $\sigma _{2}$ & line
\\ \hline
\end{tabular}%
\end{table*}

\begin{table*}[htbp]
\caption{Superconducting gap basis functions $\mathbf{d}(\mathbf{k})$ on
tetragonal lattice for odd parity, orbital symmetric and spin triplet
pairing states.}
\label{rpD4}%
\begin{tabular}{|c|c|c|c|}
\hline
$\Gamma $ & basis $\mathbf{d}(\mathbf{k})$ & $\Omega $ & gap \\ \hline\hline
& $k_{x}\hat{x}+k_{y}\hat{y};k_{z}\hat{z}$ (3D) & $\sigma _{0}$ &  \\ 
\cline{2-3}
$0$ ($A_{1u}$, $\Gamma _{1}^{-}$) & $k_{x}\hat{x}-k_{y}\hat{y}$ & $\sigma
_{3}$ & line,full \\ \cline{2-3}
& $k_{y}\hat{x}+k_{x}\hat{y}$ & $\sigma _{1}$ &  \\ \hline\hline
& $k_{y}\hat{x}-k_{x}\hat{y}$ & $\sigma _{0}$ &  \\ \cline{2-3}
$\tilde{0}$ ($A_{2u}$, $\Gamma _{2}^{-}$) & $k_{y}\hat{x}+k_{x}\hat{y}$ & $%
\sigma _{3}$ & line,full \\ \cline{2-3}
& $k_{x}\hat{x}-k_{y}\hat{y}$ & $\sigma _{1}$ &  \\ \hline\hline
& $k_{x}\hat{x}-k_{y}\hat{y}$ & $\sigma _{0}$ &  \\ \cline{2-3}
$2$ ($B_{1u}$, $\Gamma _{3}^{-}$) & $k_{x}\hat{x}+k_{y}\hat{y};k_{z}\hat{z}$
(3D) & $\sigma _{3}$ & line,full \\ \cline{2-3}
& $k_{y}\hat{x}-k_{x}\hat{y}$ & $\sigma _{1}$ &  \\ \hline\hline
& $k_{y}\hat{x}+k_{x}\hat{y}$ & $\sigma _{0}$ &  \\ \cline{2-3}
$\tilde{2}$ ($B_{2u}$, $\Gamma _{4}^{-}$) & $k_{y}\hat{x}-k_{x}\hat{y}$ & $%
\sigma _{3}$ & line,full \\ \cline{2-3}
& $k_{x}\hat{x}+k_{y}\hat{y};k_{z}\hat{z}$ (3D) & $\sigma _{1}$ &  \\ 
\hline\hline
$1$ ($E_{u}$, $\Gamma _{5}^{-}$) & $\left( k_{x}\hat{z},k_{y}\hat{z}\right)
;( k_{z}\hat{x},k_{z}\hat{y}) $ (3D) & $\sigma _{0},\sigma _{3},\sigma _{1}$
& line \\ \hline
\end{tabular}%
\end{table*}

\section{Summary and Discussions}

In summary, we have studied the pairing symmetry of the two orbital
superconducting states on a tetragonal lattice. Based on the symmetry
consideration, we have classified symmetry allowed pairing states with the
space inversion, spin, orbital, and the lattice symmetries by including a
spin-orbit coupling. In addition to the even parity for the spin singlet and
odd parity for the spin triplet pairings, familiar in the single band
superconducting gap functions, which corresponds to orbital symmetric
pairing in the two orbital systems, there are also even parity for spin
triplet and odd parity for the spin singlet pairings, corresponding to
orbital anti-symmetric pairing. The symmetry allowed gap basis functions are
listed in Tables I-IV in the text. In the orbital symmetric states, the
gap basis functions within the same representation of the point group but
with different orbital representations are allowed to combine to form a gap
function.

Below we shall discuss some limiting cases. First, we consider the weak
pairing coupling limit. In this case, we can diagonalize $H_{0}$ firstly to
obtain the two energy bands. $H_{pair}$ in Eq. (\ref{H}) is to induce a
pairing of electrons near the Fermi surfaces within a very small energy
window. If the two energy bands are not degenerate, then the two Fermi
surfaces do not coincide with each other, and the pairing will only occur
between electrons in the same band, since the energy mis-match of the two
electrons with opposite momentum in the two bands will not lead to the SC
instability in the weak coupling limit. The issue is then reduced to the two
decoupled single band problem. Because the intra-band pairing is between
symmetric orbitals, all the states with orbital anti-symmetric pairings such
as those listed in Tables II and III will not be realized. There is a
one-to-one correspondence between the present work and the single band
analysis\cite{Sigrist91}. In terms of the orbital picture, the intra-band
pairing gap function is described by a linear combination of the orbital
representations $\sigma _{0},\sigma _{1},\sigma _{3}$ in each representation
of $\Gamma $.

The strong pairing coupling case is more complicated, and possibly more
interesting. The symmetry analyses we outlined in this paper may serve as a
starting point. The pairing interaction may overcome the energy mis-match of
the paired inter-band electrons to lead to the superconductivity. In a
recent exact diagonalization calculation for a two orbital Hubbard model on
a small size system, Daghofer et al. have found an inter-orbital pairing with
spin triplet and even parity with the gap function to be $\cos {k_{x}+\cos
k_{y}}$\cite{Theory9Dagotto}. Their pairing state corresponds to $E_{g}$
representation in Table II, and provides a concrete example of the orbital
anti-symmetric pairing state. Generally we may argue that the gap structure
will be gapless with Fermi pockets for 2D systems unless the pairing coupling
is strong enough to overcome all the mis-matched paired electrons in the
momentum space. An example was given by Dai
et al.\cite{Theory1Dai} and also discussed by Wan and Wang\cite{QHWang}. This seems to
essentially rule out any possibility for line nodes in the orbital
anti-symmetric pairing state in the strong pairing coupling limit. A nodal
in quasi-particle energy requires the gap function to vanish. As a result,
the pairing strength near this nodal will not be strong enough to overcome
the energy mis-match of the inter-band paired electrons. Therefore, a nodal
in quasi-particle energy implies a Fermi pocket in this case.

Another interesting limit is the two orbitals are completely degenerate: $%
\xi_{\mathbf{k}\alpha_1,\alpha_2} =\xi_{\mathbf{k}}
\delta_{\alpha_1,\alpha_2}$. The system has an orbital SU(2) symmetry. In
this case, our analyses are most relevant, and all the classified states
listed in Tables I-IV could be stable even in the weak pairing interaction.
Because of the orientational dependence of the orbitals in crystal, such
degeneracies may not be easy to realize. A possible realization is on the
materials with two-fold pseudospin symmetry or two-valley degeneracy such as
in graphene. While the point group will depend on the precise crystal
symmetry concerned, but some general features discussed in this paper may be
applied to those systems.

We now discuss the band structure in the extended zone and the reduced zone.
Because of the positions of As atoms, the translational lattice symmetry is
reduced and the Brillouine zone is halved. In general, such a translational
symmetry reduction may lead to hopping matrix between momentum $\mathbf{k}$
and $\mathbf{k+Q}$ in the extended zone, with $\mathbf{Q}=(\pi ,\pi
)/a^{\prime }$ and $a^{\prime }=a/\sqrt{2}$ is the lattice constant of
reduced unit cell. However, for the two orbitals $d_{xz}$ and $d_{yz}$, the point
group symmetry prohibits the hybridization between states at $\mathbf{k}$
and $\mathbf{k+Q}$, if we only consider intra-layer hopping. The
tight-binding Hamiltonian adopted by both Raghu et al.\cite{Theory5Shoucheng}
and Lee and Wen\cite{Theory3Xiaogang} explicitly illustrate the vanishing
of the mixing term. Therefore, we may discuss the SC symmetry using the
extended zone and using $H_{0}$ given in Eq. (\ref{H0}). In the extended
zone, there is only one Fermi point for each $\mathbf{k}$, hence the bands
are not degenerate. In the weak pairing coupling limit, all the orbital
anti-symmetric pairing states will be irrelevant, and the weak coupling
theory will naturally lead to the orbital symmetric states.

Near the completion of the present work, we learned of the similar work by
Wan and Wang.\cite{QHWang}, who considered SC symmetry for two-orbital pairing
Hamiltonian. Our results are similar to theirs, with the difference that we
have included a spin-orbit coupling term in our group theory analysis, while
this term was not explicitly included in Ref.\cite{QHWang}. As a result, our
classification for the spin triplet states is not the same as theirs. Such
difference may be amplified when we discuss some behaviors related to spin
degrees of freedom. We also note that similar group theory analysis were
carried out for the two band pairing Hamiltonian by Wang et al.\cite{Dai2}.
Since they adopted the bands instead of the orbitals, a direct comparison is
not apparent. We become aware of another related work\cite{JRShi} after
completing the present work too.

\section{Acknowledgement}

We thank T. K. Ng and X. Dai for useful discussions and HKSAR RGC
grants for partial financial support.

\appendix

\section{The symmetry of $\xi _{\mathbf{k\protect\alpha }_{1}\alpha _{2}}$ 
in Equation (\ref{H0}) \label{xi}}

In this appendix, we will discuss the symmetry requirement of $\xi _{\mathbf{%
k\alpha }_{1}\alpha _{2}}$. The non-interacting Hamiltonian given by Eq.(\ref%
{H0}) should keep invariant under any symmetry transformation of point-group 
$D_{4}$, hence $H_{0}$ belongs to the representation $A_{1}$. This symmetry
Requirement will affect the choice of $\xi _{\mathbf{k\alpha }_{1}\alpha
_{2}}$. For convenience, we use the $2\times 2$ matrix form $\hat{\xi}_{%
\mathbf{k}}$ in orbital space, thus $\hat{\xi}_{\mathbf{k}}$ can be
rewritten in terms of Pauli matrices, $\hat{\xi}_{\mathbf{k}}=\xi _{\mathbf{k%
}}^{0}\sigma _{0}+\xi _{\mathbf{k}}^{1}\sigma _{1}+\xi _{\mathbf{k}%
}^{2}\sigma _{2}+\xi _{\mathbf{k}}^{3}\sigma _{3}$. Similarly to the case of 
$\Omega $, $\phi _{\mathbf{k}s}^{\dag }\sigma _{0}\phi _{\mathbf{k}s}$, $%
\phi _{\mathbf{k}s}^{\dag }\sigma _{1}\phi _{\mathbf{k}s}$, $\phi _{\mathbf{k%
}s}^{\dag }\sigma _{2}\phi _{\mathbf{k}s}$ and $\phi _{\mathbf{k}s}^{\dag
}\sigma _{3}\phi _{\mathbf{k}s}$ transform as $A_{1}$, $B_{2}$, $A_{2}$ and $%
B_{1}$ respectively, where $\phi _{\mathbf{k}s}=(c_{\mathbf{k}1},c_{\mathbf{k%
}2})^{T}$. Using the CG coefficients of $D_{4}-C_{4}$ group chain\cite%
{Butler}, we find that $\xi _{\mathbf{k}}^{0}$, $\xi _{\mathbf{k}}^{1}$, $%
\xi _{\mathbf{k}}^{2}$ and $\xi _{\mathbf{k}}^{3}$ transform as $A_{1}$, $%
B_{2}$, $A_{2}$ and $B_{1}$ respectively. Some examples of $\xi _{\mathbf{k}%
}^{0,1,2,3}$ are shown in the following, 
\begin{eqnarray*}
\xi _{\mathbf{k}}^{0} &=&1,\cos k_{x}+\cos k_{y},\cos k_{x}\cos k_{y}, \\
\xi _{\mathbf{k}}^{1} &=&\sin k_{x}\sin k_{y}, \\
\xi _{\mathbf{k}}^{2} &=&\sin k_{x}\sin k_{y}\left( \cos k_{x}-\cos
k_{y}\right) , \\
\xi _{\mathbf{k}}^{3} &=&\cos k_{x}-\cos k_{y}.
\end{eqnarray*}

\section{The point group $D_{4h}$ \label{gD4h}}

\begin{figure}[htpb]
\includegraphics[width=7.0cm]{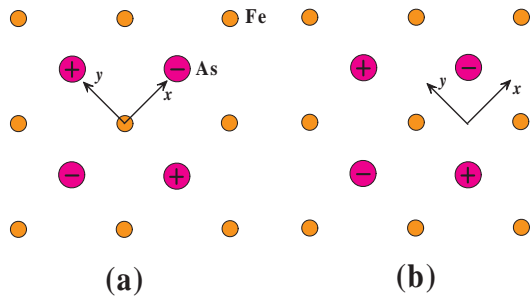}
\caption{(color online) FeAs layer and specification of $(x,y,z)$. $xy$ plane consisting of
Fe atoms is shown in (a) and (b) with different origin choices. $z$ axis is
perpendicular to Fe-plane. \textquotedblleft$+$\textquotedblright denotes 
an As atom above the Fe-plane, while \textquotedblleft$-$\textquotedblright 
denotes an As atom below the Fe-plane. (a) Origin choice 1. (b) Origin
choice 2. (also see Fig.\ref{lat}) }
\label{FeAs}
\end{figure}

Here we would like to specify the tetragonal point group according to the 
$LaFeAsO$ space group $P4/nmm$.\cite{crystallography} In real space, the
point group is neither usual $D_{4h}=D_{4}\otimes \sigma _{h}$ nor usual $%
D_{4}$ generated by $\left\{ C_{4z},C_{2y}\right\} $, where $\sigma _{h}$ is
the reflection refer to $xy$ plane, $C_{4z}$ is the four-fold rotation
around the $z$ axis, and $C_{2y}$ is the two-fold rotation around the $y$ axis,
where $(x,y,z)$ is specified in Fig. \ref{FeAs}. However, it contains two
subgroups, which refer to two different origin choices of the lattice. One is a
subgroup of $D_{4h}$ generated by $\left\{ C_{4z}\sigma _{h},C_{2y}\sigma
_{h}\right\} $, which is also a $D_{4}$ group (or to be precise, $D_{2d}$,
an isomorphic group to $D_{4}$) with origin choice 1 as shown in Figs. \ref%
{lat}(a) and \ref{FeAs}(a). The other subgroup is the direct product of
inversion symmetry group $I$ and cyclic group $C_{2xy}$ with origin choice 2
as shown in Figs. \ref{lat}(b) and \ref{FeAs}(b). The transformation of $%
\left( x,y,z\right) $ under these symmetry operations can be found in Tables %
\ref{D4h1} and \ref{D4h2}. Hence in $\mathbf{k}$-space, it is still a
tetragonal point group $D_{4h}$.

\begin{table}[htbp]
\caption{Eight symmetry operations of $D_{2d}$ (an isomorphic group to $%
D_{4} $) will generate eight general positions. Where \textquotedblleft
general\textquotedblright\ is defined as the following: a set of symmetrical
equivalent points is said to be in \textquotedblleft general
position\textquotedblright\ if each of its points is left invariant only by
the identity operation but by no other symmetry operation of the space
group. The origin choice is 1.}
\label{D4h1}%
\begin{tabular}{|c|c|}
\hline
group element & general position \\ \hline
$E$ & $\left( x,y,z\right) $ \\ \hline
$C_{4z}\sigma _{h}$ & $\left( y,-x,-z\right) $ \\ \hline
$\left( C_{4z}\sigma _{h}\right) ^{2}$ & $\left( -x,-y,z\right) $ \\ \hline
$\left( C_{4z}\sigma _{h}\right) ^{3}$ & $\left( -y,x,-z\right) $ \\ \hline
$C_{2y}\sigma _{h}$ & $\left( -x,y,z\right) $ \\ \hline
$C_{2y}\sigma _{h}C_{4z}\sigma _{h}$ & $\left( -y,-x,-z\right) $ \\ \hline
$C_{2y}\sigma _{h}\left( C_{4z}\sigma _{h}\right) ^{2}$ & $\left(
x,-y,z\right) $ \\ \hline
$C_{2y}\sigma _{h}\left( C_{4z}\sigma _{h}\right) ^{3}$ & $\left(
y,x,-z\right) $ \\ \hline
\end{tabular}%
\end{table}

\begin{table}[htbp]
\caption{Four symmetry operations and corresponding general positions of
group $I\times C_{2xy}$. The origin choice is 2.}
\label{D4h2}%
\begin{tabular}{|c|c|}
\hline
group element & general position \\ \hline
$E$ & $\left( x,y,z\right) $ \\ \hline
$C_{i}$ & $\left( -x,-y,-z\right) $ \\ \hline
$C_{2xy}$ & $\left( y,x,z\right) $ \\ \hline
$C_{i}C_{2xy}$ & $\left( -y,-x,-z\right) $ \\ \hline
\end{tabular}%
\end{table}

There are five irreducible representations of $D_{4}$ group, four of them, $%
A_{1}$, $A_{2}$, $B_{1}$ and $B_{2}$ are one-dimensional representations,
and one of them, $E$ is a two-dimensional representation. All these five
representations are representations of group $D_{4h}$ too. However, there
are 2 two-dimensional irreducible representations $E^{\prime }$ and $%
E^{\prime \prime }$ of $D_{4h}$ group, neither of them is the representation
of group $D_{4}$. Naively, $E^{\prime }$ and $E^{\prime \prime }$ can be
viewed as subrepresentations of two irreducible representations of group $SU(2)$, 
$J=1/2$ and $J=3/2$, respectively. Since the representations $
E^{\prime }$ and $E^{\prime \prime }$ can not result in quadratic terms in
Hamiltonian or Ginzburg-Landau free energy, we will not discuss them in this
paper.

\section{Transformation of gap functions \label{transfer}}

It is not $\Delta \left( \mathbf{k}\right) $ but $\psi \left( \mathbf{k}%
\right) $ and $\mathbf{d}\left( \mathbf{k}\right) $ transform as
representations of symmetry group. In this appendix, we list the
transformations of $\psi \left( \mathbf{k}\right) $ and $\mathbf{d}\left( 
\mathbf{k}\right) $ under various symmetry operations. Firstly, under a
point-group transformation $g$, $\psi \left( \mathbf{k}\right) $ and $%
\mathbf{d}\left( \mathbf{k}\right) $ transform as%
\begin{eqnarray}
g\psi \left( \mathbf{k}\right) &=&\psi \left( D_{\left( G\right) }^{-}\left(
g\right) \mathbf{k}\right) ,  \notag \\
g\mathbf{d}\left( \mathbf{k}\right) &=&D_{\left( G\right) }^{+}\left(
g\right) \mathbf{d}\left( D_{\left( G\right) }^{-}\left( g\right) \mathbf{k}%
\right) ,
\end{eqnarray}%
where $D_{\left( G\right) }^{\pm }\left( g\right) $ is the representation in
three-dimensional space with positive (spin-space) or negative ($\mathbf{k}$%
-space) respectively. Secondly, time-reversal transformations of $\psi
\left( \mathbf{k}\right) $ and $\mathbf{d}\left( \mathbf{k}\right) $ take
the forms,%
\begin{equation}
K\psi \left( \mathbf{k}\right) =\psi ^{\ast }\left( -\mathbf{k}\right) ,K%
\mathbf{d}\left( \mathbf{k}\right) =-\mathbf{d}^{\ast }\left( -\mathbf{k}%
\right) .  \label{K1}
\end{equation}%
The anti-symmetric nature of Fermion systems, see Eq. (\ref{fasym}), will lead to%
\begin{equation}
\psi \left( -\mathbf{k}\right) =\psi \left( \mathbf{k}\right) ,\mathbf{d}%
\left( -\mathbf{k}\right) =-\mathbf{d}\left( \mathbf{k}\right) ,  \label{syd}
\end{equation}%
for symmetric $\Omega $ and%
\begin{equation}
\psi \left( -\mathbf{k}\right) =-\psi \left( \mathbf{k}\right) ,\mathbf{d}%
\left( -\mathbf{k}\right) =\mathbf{d}\left( \mathbf{k}\right) ,  \label{asyd}
\end{equation}%
for anti-symmetric $\Omega $. Hence, combining the above and the Hermitian
choice of $\Omega $'s, the time-reversal invariance conditions for $\psi
\left( \mathbf{k}\right) $ and $\mathbf{d}\left( \mathbf{k}\right) $ become%
\begin{equation}
\psi ^{\ast }\left( \mathbf{k}\right) =\psi \left( \mathbf{k}\right) ,%
\mathbf{d}^{\ast }\left( \mathbf{k}\right) =\mathbf{d}\left( \mathbf{k}%
\right) ,  \label{K2}
\end{equation}%
since under time-reversal transformation, $\Omega $ transforms as%
\begin{equation}
K\Omega =\Omega ^{\ast }.  \label{K3}
\end{equation}

\section{Energy gap functions in the degenerate bands \label{gap}}

The energy gap of the superconducting states indeed depends on the details
of interaction, especially, depends on the ratio of $\delta t/\lambda $,
where $\delta t$ is the energy scale of the splitting of two bands and $%
\lambda $ is the energy scale of pairing potential. In the the
\textquotedblleft strong pairing coupling\textquotedblright\ limit $\delta t\ll
\lambda $, we expect the energy gap is close to $\delta t=0$ case, say, two
bands are degenerate. A small perturbation proportional to $\delta t/\lambda 
$ will not change the energy gap very much, e.g. close the full gap or
change from full gap to line nodal gap. In the weak pairing coupling limit $\lambda
\ll \delta t$, the situation may be very different from strong coupling
limit, which is discussed in Ref.\cite{QHWang}. So that we will focus on the
strong coupling limit and assume two degenerate bands in the following.

Due to two degenerate bands, the effective mean field Hamiltonian in $%
\mathbf{k}$-space can be written as an $8\times 8$ matrix,%
\begin{equation}
\hat{H}_{\mathbf{k}}=\left( 
\begin{array}{cc}
\xi _{\mathbf{k}}\sigma _{0}\otimes \sigma _{0} & \Delta (\mathbf{k}) \\ 
\Delta ^{\dag }(\mathbf{k}) & -\xi _{\mathbf{k}}\sigma _{0}\otimes \sigma
_{0}%
\end{array}%
\right) ,
\end{equation}%
with the basis $\mathbf{c}_{\mathbf{k}}=(c_{\mathbf{k}\uparrow 1},c_{\mathbf{%
k}\uparrow 2},c_{\mathbf{k}\downarrow 1},c_{\mathbf{k}\downarrow 2},c_{-%
\mathbf{k}\uparrow 1}^{\dag },c_{-\mathbf{k}\uparrow 2}^{\dag },c_{-\mathbf{k%
}\downarrow 1}^{\dag },c_{-\mathbf{k}\downarrow 2}^{\dag })^{T}$. The
indices in the $4\times 4$ matrices $\Delta (\mathbf{k})$ and $\xi _{\mathbf{%
k}}\sigma _{0}\otimes \sigma _{0}$ are arranged as the following, by direct
products the former two indices denote spin space and the later two denote
two orbitals. It is easy to know the energy dispersion, 
\begin{equation}
E_{\mathbf{k}\mu }=\pm \sqrt{\xi _{\mathbf{k}}^{2}+\Delta _{\mathbf{k}\mu
}^{2}},
\end{equation}%
where $\Delta _{\mathbf{k}\mu }^{2}$ is one of the eigenvalues of the matrix 
$\Delta (\mathbf{k})\Delta ^{\dag }(\mathbf{k})$. For degenerate bands, the
minimum of $\left\vert \Delta _{\mathbf{k}\mu }\right\vert $ is the energy
gap. For simplicity, we will focus on the $k_{z}$-independent pairing with a
cylinder-like Fermi surface which is the case of doped $LaFeAsO$ most likely.

At first, we will consider the even parity, orbital anti-symmetric, spin
triplet pairing states in Table II. Orbital anti-symmetric states have only
one component $\sigma _{2}$ in the $\Omega $ part. Gap function is of the
form, $\Delta (\mathbf{k})=i\left[ \mathbf{\sigma }\cdot \mathbf{d}\left( 
\mathbf{k}\right) \right] \sigma _{2}\otimes \sigma _{2}$, thus $\Delta _{%
\mathbf{k}\mu }^{2}=\left\vert \mathbf{d}\right\vert ^{2}\pm \left\vert 
\mathbf{d\times d}^{\ast }\right\vert $. For the time-reversal invariant
state, $\mathbf{d}=\mathbf{d}^{\ast }$, the gapless condition follows as $%
\left\vert \mathbf{d}\right\vert ^{2}=0$. For $B_{1g}$ and $B_{2g}$ states,
they are $d$-wave states and have line nodal gap at Fermi surfaces. $A_{1g}$
states can be of either $s$-wave or extended $s$-wave. The $s$-wave state is
of full gap while the extended $s$-wave state possibly has line nodal gap at
Fermi surface, e.g., the state $\mathbf{d}\left( \mathbf{k}\right) =\cos
k_{x}\cos k_{y}\hat{z}$. The $E_{g}$ representation involves $s$-wave,
extended $s$-wave and $d$-wave states. The $s$-wave state is fully gapful, the 
$d$-wave state has line nodal gap, the extended $s$-wave state can be either
fully gapful or of line nodal gap.

Then we consider the odd parity, orbital symmetric, spin triplet pairing
states in Table IV. Orbital symmetric states have three components $\sigma
_{0,1,3}$ in the $\Omega $ part. Gap function can be written as 
\begin{widetext}
\begin{equation}
\Delta (\mathbf{k})=i\left[ \mathbf{\sigma }\cdot \mathbf{d}_{0}\left( 
\mathbf{k}\right) \right] \sigma _{2}\otimes \sigma _{0}+i\left[ \mathbf{%
\sigma }\cdot \mathbf{d}_{1}\left( \mathbf{k}\right) \right] \sigma
_{2}\otimes \sigma _{1}+i\left[ \mathbf{\sigma }\cdot \mathbf{d}_{3}\left( 
\mathbf{k}\right) \right] \sigma _{2}\otimes \sigma _{3},
\end{equation}%
thus%
\begin{eqnarray}
\Delta (\mathbf{k})\Delta ^{\dag }(\mathbf{k}) &=&[ ( \left\vert 
\mathbf{d}_{0}\right\vert ^{2}+\left\vert \mathbf{d}_{1}\right\vert
^{2}+\left\vert \mathbf{d}_{3}\right\vert ^{2}) \sigma _{0}+i\left( 
\mathbf{d}_{0}\mathbf{\times d}_{0}^{\ast }+\mathbf{d}_{1}\mathbf{\times d}%
_{1}^{\ast }+\mathbf{d}_{3}\mathbf{\times d}_{3}^{\ast }\right) \cdot 
\mathbf{\sigma }] \otimes \sigma _{0}  \notag \\
&&+\left[ \left( \mathbf{d}_{0}\cdot \mathbf{d}_{1}^{\ast }+\mathbf{d}%
_{1}\cdot \mathbf{d}_{0}^{\ast }\right) \sigma _{0}+i\left( \mathbf{d}_{0}%
\mathbf{\times d}_{1}^{\ast }+\mathbf{d}_{1}\mathbf{\times d}_{0}^{\ast
}\right) \cdot \mathbf{\sigma }\right] \otimes \sigma _{1}  \notag \\
&&+\left[ \left( \mathbf{d}_{0}\cdot \mathbf{d}_{3}^{\ast }+\mathbf{d}%
_{3}\cdot \mathbf{d}_{0}^{\ast }\right) \sigma _{0}+i\left( \mathbf{d}_{0}%
\mathbf{\times d}_{3}^{\ast }+\mathbf{d}_{3}\mathbf{\times d}_{0}^{\ast
}\right) \cdot \mathbf{\sigma }\right] \otimes \sigma _{3}  \notag \\
&&+\left[ \left( \mathbf{d}_{1}\mathbf{\times d}_{3}^{\ast }-\mathbf{d}_{3}%
\mathbf{\times d}_{1}^{\ast }\right) \cdot \mathbf{\sigma }-i\left( \mathbf{d%
}_{1}\cdot \mathbf{d}_{3}^{\ast }-\mathbf{d}_{3}\cdot \mathbf{d}_{1}^{\ast
}\right) \sigma _{0}\right] \otimes \sigma _{2}.
\end{eqnarray}%
For a time-reversal invariant state, $\mathbf{d}_{i}^{\ast }\left( \mathbf{k}%
\right) =\mathbf{d}_{i}\left( \mathbf{k}\right) $, $i=0,1,3$, so that the
above can be simplified as%
\begin{equation}
\Delta (\mathbf{k})\Delta ^{\dag }(\mathbf{k})=\left( \mathbf{d}_{0}^{2}+%
\mathbf{d}_{1}^{2}+\mathbf{d}_{3}^{2}\right) \sigma _{0}\otimes \sigma
_{0}+2\left( \mathbf{d}_{0}\cdot \mathbf{d}_{1}\right) \sigma _{0}\otimes
\sigma _{1}+2\left( \mathbf{d}_{0}\cdot \mathbf{d}_{3}\right) \sigma
_{0}\otimes \sigma _{3}+2\left( \mathbf{d}_{1}\mathbf{\times d}_{3}\right)
\cdot \mathbf{\sigma }\otimes \sigma _{2}.
\end{equation}%
We obtain from the above
\begin{equation}
\Delta _{\mathbf{k}\mu }^{2}=\left( \mathbf{d}_{0}^{2}+\mathbf{d}_{1}^{2}+%
\mathbf{d}_{3}^{2}\right) \pm 2\sqrt{\left( \mathbf{d}_{0}\cdot \mathbf{d}%
_{1}\right) ^{2}+\left( \mathbf{d}_{0}\cdot \mathbf{d}_{3}\right)
^{2}+\left( \mathbf{d}_{1}\mathbf{\times d}_{3}\right) ^{2}}.
\end{equation}%
Gapless condition reads%
\begin{equation}
\mathbf{d}_{0}^{2}+\mathbf{d}_{1}^{2}+\mathbf{d}_{3}^{2}=2\sqrt{\left( 
\mathbf{d}_{0}\cdot \mathbf{d}_{1}\right) ^{2}+\left( \mathbf{d}_{0}\cdot 
\mathbf{d}_{3}\right) ^{2}+\left( \mathbf{d}_{1}\mathbf{\times d}_{3}\right)
^{2}}.  \label{gapt}
\end{equation}%
\end{widetext}
Careful analysis shows node can appear only when at least one of $\left\vert 
\mathbf{d}_{0}\right\vert $, $\left\vert \mathbf{d}_{1}\right\vert $ and $%
\left\vert \mathbf{d}_{3}\right\vert $ vanish. So that $E_{u}$ states in
Table IV are of line nodal gap. The other four representations $A_{1u}$, $%
A_{2u}$, $B_{1u}$ and $B_{2u}$ can be of either line nodal or full gap. For
example, for an $A_{1u}$ states in Table IV which consists of two components
in the $\Omega $ part, $\mathbf{d}_{0}\left( \mathbf{k}\right) =\sin k_{x}%
\hat{x}+\sin k_{y}\hat{y}$, $\mathbf{d}_{3}\left( \mathbf{k}\right) =\sin
k_{x}\hat{x}-\sin k_{y}\hat{y}$ and $\mathbf{d}_{1}\left( \mathbf{k}\right)
=0$, nodal lines will appear at $\sin k_{x}=0$ and $\sin k_{y}=0$. Moreover, 
any $A_{1u}$ state in Table IV which consists of only one component in the $%
\Omega $ part is of full gap.

Similar consideration will lead to the results for spin singlet states shown
in Tables I and III. Of course, when the ratio $\delta t/\lambda $
becomes large, the situation may change. This change strongly depends on the
details of both pairing states and the Hamiltonian. For example, for an $s$%
-wave with $\mathbf{d}\left( \mathbf{k}\right) =\hat{z}$, and $\Omega
=\sigma _{2}$, Fermi pockets may appear when $\delta t$ and $\lambda $ are
comparable.


\begin{thebibliography}{99}
\bibitem{P4nmm} If the lattice undergoes a structural transition,
e.g., the transition from tetragonal ($P4/nmm$) to monoclinic ($P112/n$) 
in the parent compound $LaFeAsO$, lattice symmetry will be reduced 
and thereby the pairing states within 
each representation of $D_4$ will be further splitted by energy. 
However, $P_{space}$, $P_{spin}$ and $P_{orbital}$
considered in this paper are still good quantum numbers. Moreover, when
the monoclinic distortion to tetragonal phase is small as in practice,\cite{Neutron}
our study is a good approximation to such a monoclinic lattice.
\end{thebibliography}
\end{document}